\documentclass[pra,onecolumn,superscriptaddress]{revtex4}%
\usepackage{amssymb}
\usepackage{amsmath}
\usepackage{graphicx}
\usepackage{dcolumn}
\usepackage{xcolor}
\usepackage{bm}
\usepackage{subfigure}
\usepackage{amsfonts}
\usepackage{appendix}
\usepackage{tikz}
\usetikzlibrary{quantikz2,backgrounds,fit,decorations.pathreplacing}
\usepackage{xcolor}%
\providecommand{\U}[1]{\protect\rule{.1in}{.1in}}
\begin{document}
\preprint{APS/123-QED}
\title{ Real-time diagnostics  on a  QKD link via
QBER Time Series Analysis }
\author{Georgios Maragkopoulos}

\affiliation{Department of Informatics and Telecommunications, National and Kapodistrian
University of Athens, Panepistimiopolis, Ilisia, 15784, Greece}

\author{ Aikaterini Mandilara}

\affiliation{Department of Informatics and Telecommunications, National and Kapodistrian
University of Athens, Panepistimiopolis, Ilisia, 15784, Greece}
\affiliation{Eulambia Advanced Technologies, Agiou Ioannou 24, Building Complex C, Ag. Paraskevi, 15342, Greece}
\author{ Thomas Nikas}

\affiliation{Department of Informatics and Telecommunications, National and Kapodistrian
University of Athens, Panepistimiopolis, Ilisia, 15784, Greece}

\author{ Dimitris Syvridis}

\affiliation{Department of Informatics and Telecommunications, National and Kapodistrian
University of Athens, Panepistimiopolis, Ilisia, 15784, Greece}
\affiliation{Eulambia Advanced Technologies, Agiou Ioannou 24, Building Complex C, Ag. Paraskevi, 15342, Greece}

\begin{abstract}
The integration of QKD systems in Metro optical  networks raises   challenges which cannot be completely resolved with the current technological status. In this work we devise a methodology for identifying different kind of impairments which may occur on the quantum channel during its transmission in an operational network. The methodology is built around a supervised ML pipeline which is using as input QBER and SKR time-series and requires no further interventions on the QKD system. The identification of impairments happens in real time and even though such information cannot reverse incidents, this can be valuable for users, operators and   key management system.
\end{abstract}
\maketitle
\section{Introduction}

After three decades of intense research \cite{In1} in quantum cryptographic protocols and their implementations,  we are in the era where  Quantum Key Distribution (QKD) systems are commercially available and ready for use in the labs or  in real life. While  technology in such devices is very advanced and keeps improving,    there is  a couple of important obstacles  which prevent their generalized integration in existing Metro optical networks.
The first obstacle is the vulnerability of quantum signals
to the effect of attenuation due to propagation in  fiber or due to the presence of network components. A quantum signal for QKD purposes contains, in average, less than one photon
and therefore a simple incident of a photon loss is  destructive for the  carried quantum information.
Quantum repeaters \cite{In2}, error correcting codes \cite{In3}, twin-field protocols \cite{In4} are designed to solve in principle this problem but the technology is still immature for their generalized application.
The second important obstacle is the effect of photons addition to quantum signal in the case where this coexists   with  classical ones propagating in the same optical fiber. This effect  in large extent restricts QKD signals to   Single-mode Dark Fibers (SDF) and numerous studies have been carried out  on the performance's degradation of QKD systems under different  conditions of coexistence dependent on the wavelength spacing between classical and quantum  channels,  optical power of the former, etc.
 
 In this work  we seek for solutions  offered by the field of Machine Learning (ML) for 
advancing the integration of  QKD devices in classical communication networks taking into account the aforementioned obstacles. More specifically, we develop an ML methodology to classify different kinds of
impairments (producing irreversible types of errors) which may occur on a quantum channel  assuming only  access to  QBER and Secure Key Rate (SKR) data. This work follows the latest trend of improving the functionality of QKD  via ML methods \cite{MLQKD1, MLQKD2, MLQKD3, MLQKD4, MLQKD5, MLQKD6, MLQKD7,  MLQKD10}.   The  objective of the reported work is to  provide tools for performing real time diagnostics of the QKD system and to extract information on  the impairments affecting the operation of the system. Such information can be particularly useful for the Key Management System (KMS) of the network which controls the `flaw' of keys.

The structure of the manuscript is as follows. We first present an optimized ML methodology for  extracting features from QBER/SKR time series and performing classification according to these. In order to test the validity of the method we perform experiments to obtain training and test data under  conditions which experimentally emulate different types of impairments. We  present the results of classification for our experimental data  and finally we discuss on the key-points and  perspectives of the method.

\section{ML methods}
\begin{figure*}[t]
    \centering
    \includegraphics[height=3cm]{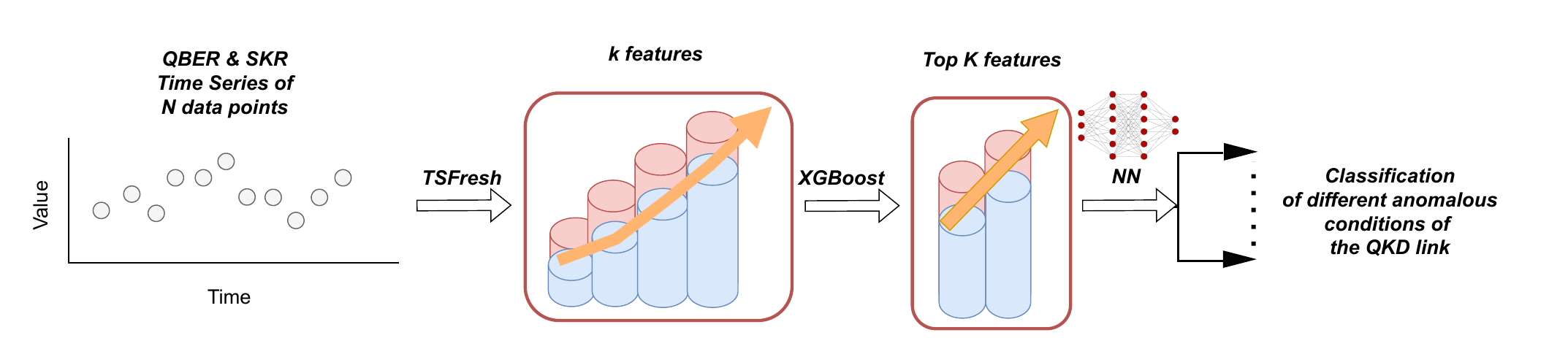}  
  \caption{ \label{fig0} While  QKD system is functioning 
and distributes keys, one acquires $N$ sequential   values of QBER and SKR. $k$ features are extracted from these time-series with  \textit{tsfresh} python package, which in sequence are reduced to $K$ with \textit{xgboost} method.  The $K$ values are fed in a deep  NN for either training (during the training phase) or class prediction (during prediction phase).  }
\end{figure*}

Different  QKD systems  differ in the  QKD protocols which these implement as well as in the error correcting and authentication methods.  On the other hand all QKD devices    provide SKR and  QBER data. We set up an ML `pipeline', sketched in Fig.~\ref{fig0}.,  that  is fed with  QBER and SKR data from a QKD device.
The parameters inside are first trained with normal data and data in anomalous conditions to classify the different working phases of the 
QKD link. Then by feeding this pipeline with real time sequential values of QBER and SKR from the device, one can classify the current status of the QKD link.

In more details, we use \textit{tsfresh} python package \cite{MLQKD9} to extract features for each batch of $N$  QBER and SKR data points. In our studies where we set $N=10$,
 the total number $k$ of extracted features for each batch  is greater than $1500$. These  include both simple statistical measures such as mean, variance, quantiles, and more complex ones such as  \textit{arima} coefficients, \textit{Fourier} and wavelet transformations. A feature space of this size increases the chances of over-fitting, as well as training and inference times. For these reasons, a \textit{xgboost} model \cite{MLQKD8} is trained in order to reduce the \textit{feature} space. This is achieved by choosing the top $K$ in terms of information gain. The choice of xgboost  in our studies has proven advantageous to  more standard methods such as PCA mainly due to  sparseness and  missing values in the feature space. In our applications of the method, we pick the $K=50$ top performers provided by xgboost which 
 are then fed as input in a Neural Network (NN) whose architecture has a depth of 3 hidden layers  (50x128x256x128x9) and where cross-entropy loss is used in the  training phase.

Let us now describe the data-acquisition in training and prediction phases with a real-time usage example of the method.\\
\textbf{Training }
    \begin{itemize}
    \item Activate the QKD system and   draw the first $N$ data points. These data points are used as
       reference points and all next points/time series should be normalized according to the median values of these data, using
       MinMax scaler.
       \item Proceed by acquiring sequential log files with N values of QBER and SKR while
       inducing impairments on the transmission line of the quantum signal. The data are then  labeled  according to the type of impairment.
       \item Use all collected data to train together the xgboost model and NN so that  the data  are classified to different labels
      while the cross-entropy loss is minimized. 
     \end{itemize}
     \textbf{Prediction }
    \begin{itemize}
    \item As for the training phase,  we assume a period where the
    system runs without impairment and collect $N$ data points as reference ones. 
   \item At every time step, one may feed the batch created by the current point and $N-1$ previous ones in the ML pipeline in order to conclude on impairments present  on the QKD link.
     \end{itemize}

\section{Demonstration of the method with experimental data}

The core of the experimental setting is a pair of Toshiba terminals, QKD4.2A-MU  and QKD4.2B-MU,   realizing an  advanced one-way phase-encoded protocol with coherent states, the so called T12  protocol \cite{T12}.  QKD Transmitter (Alice) is sending quantum data at $~1310$ nm to the Receiver (Bob) via  a SDF.  In the same fiber  two  low-power classical signals are co-propagating  at $1530$ and $1529.30$ nm.   A second auxiliary fiber connects the terminals for establishing classical communication from  Receiver to Transmitter at $1528.77$ nm. The power of all three classical channels is $< +3$ dBm.

\begin{figure*}[t]
    \centering
    \includegraphics[height=5cm]{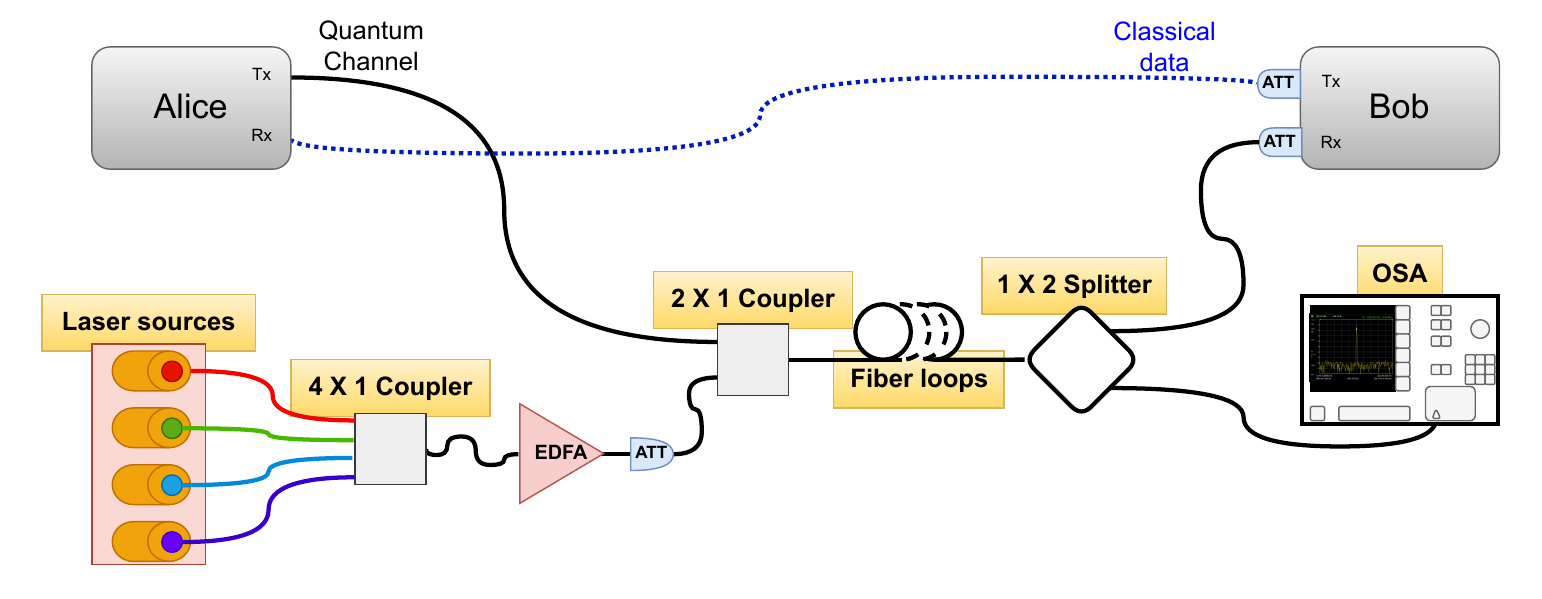}  
  \caption{ \label{fig1} Experimental setting for QBER/SKR data acquisition. From the bank of lasers one can select the  number, $1-4$, and power of classical signals placed at the C-band coexisting with quantum data placed in the O-band. An Erbium Doped Fiber Amplifier (EDFA) can be connected after the laser bank to further amplify the power of lasers.  A number of fiber loops of diameter $\approx 2$ cm can be formed along the QKD link to induce photon losses.  An Optical Spectrum Analyzer  (OSA) is used to monitor the attenuation on the fiber link. }
\end{figure*}

The overall experimental setting for training-test data acquisition under different conditions which mark the different classes of impairments, is presented in Fig.~\ref{fig1}. One can use this setting to realize  $4$ different configurations for acquiring data: a) Normal mode, i.e., no impairments along the transmission line, b) Coexistence with optical cw signals without amplification via EDFA, c)  Coexistence with optical cw signals amplified by the EDFA, d) Photon loss induced by small fiber loops. The attenuation along the transmission line due to fiber, and optical elements (couplers, attenuators) is  for all experiments at $-14$ dB.  Below, we give quantitative details for the categorical classes of Tab.~\ref{tab:mytable2} namely the wavelength $\lambda$ and  power $P$ of lasers (as  measured by  OSA, see Fig.~\ref{fig1}), as well as the excess attenuation $A_{exc}$ due to the presence of fiber's loops.
In the training phase of coexistence with optical cw signals without EDFA we have
collected data for an interval of powers as indicated below and have treated all these data under the same class/label. We have done so in order to investigate whether the model is capable to get trained for a wide class of events.
\begin{itemize}
\item \textbf{Class 1.} 1 Laser:   $\lambda= 1549.38 $ nm, $P=-\left\{23.5, 21.7, 20.5, 19.55, 18.84,18.37, 18.1\right\}$ dBm. 
\item \textbf{Class 2.} 2 Lasers:
$\lambda_1= 1549.38$, $\lambda_2=1549.46$ nm, \\ $P_1=-\left\{23.5, 21.7,20.5, 19.55, 18.84, 18.37, 18.1\right\}$, \\ $P_2=-\left\{21.6, 20.2, 19.4, 19.0, 18.8, 18.9, 19.2\right\}$ dBm.
\item 4 Lasers $\&$ EDFA:
$\lambda_1= 1548.5$, $\lambda_2=1549$, $\lambda_3= 1549.5$ , $\lambda_4=1550$ nm, \\
\textbf{Class 3.} $I=18$ mA: $P_1=-17.9$,
$P_2=-16.9$,  $P_3=-15.6$,
$P_4=-15.6$ dBm.\\
\textbf{Class 4.} $I=21$ mA: $P_1=-16.5$,
$P_2=-15.7$,  $P_3=-14.6$,
$P_4=-14.3$ dBm.\\
\textbf{Class 5.} $I=24$ mA: $P_1=-15.5$,
$P_2=-14.5$,  $P_3=-13.4$,
$P_4=-13.1$ dBm.
\item \textbf{Class 6.} Photon Loss 20\%: $A_{exc}=-0.9$ dB.
\item \textbf{Class 7.} Photon Loss 46\%: $A_{exc}=-1.9$ dB.
\item \textbf{Class 8.} Photon Loss 67\%: $A_{exc}=-3.1$ dB.
\end{itemize}

In the experiments we make distinction between lasers with and without EDFA since   this type of amplifier incorporates an optical filter which decreases the outband ASE noise leaking to O band where the quantum channel resides.  Regarding the experiments where we simulate the effect of photon losses via the formation of small loops. In these formations  the radius of curvature of fiber's coating is decreased,  impelling a part of the quantum light signal, proportional to the number of loops,  to  escape out of the fiber. This experiment emulates in a controllable way  the action of an eavesdropper or the activation of an optical element, such as coupler  or multiplexer in the network.  

\begin{figure*}[h]
    \centering
    \includegraphics[height=12cm]{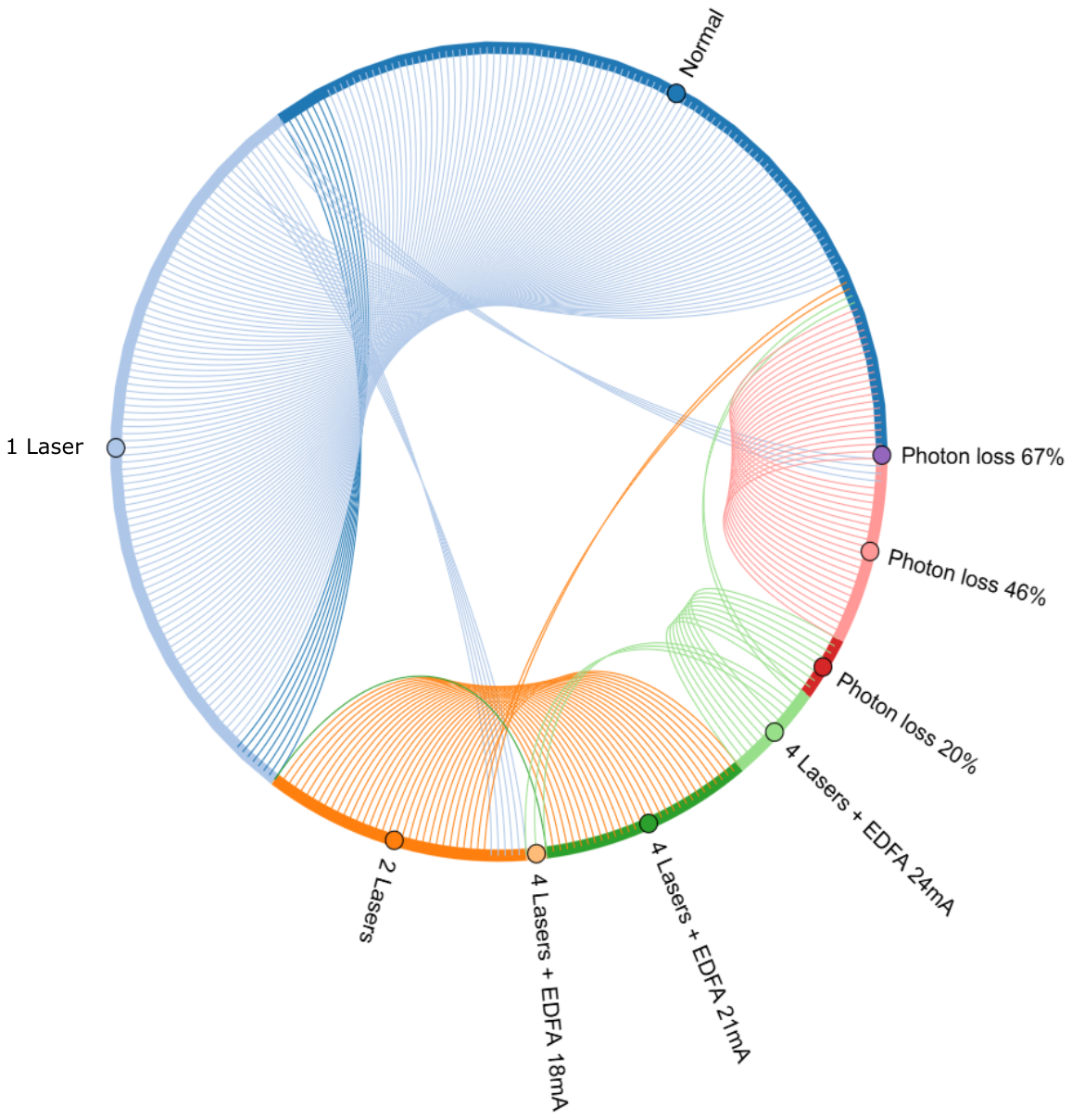}  
  \caption{ \label{fig2} This chord diagram \cite{miscla} visualizes the misclassification events for the test set. Each colored segment of the circle
represents a class and the related arc length is proportional to the percentage of misclassified data for this class.  
The colored stripes indicate misclassification of data from one class
to another, having the color of the former. As an example, an  orange stripe connects the orange class (2
Lasers) to dark green class (4 Lasers + EDFA 21mA). This stripes indicates that  datapoints with 
ground-truth label  `2 Lasers' are misclassified as `4 Lasers + EDFA 21mA'.}
\end{figure*}

We proceed with the application of ML methodology on the acquired data and we first  split these as 80\%-20\% for training-test respectively. The test data are always derived from the tail end of the time series since a plain random split would result to a training set which contains data from time frames ahead of some test data, that could lead to misleading results. After training the xgboost  and NN models, predictions were made on the test set and the results are provided in Tab.~\ref{tab:mytable2}. These are also complemented with the chord  diagram of Fig.~\ref{fig2} that intends to give a visual representation of the misclassified data.
The results of the table show that the \textit{Precision}, i.e., the number of true positives divided by the total number of positive predictions, is high for all impairment types. On the other hand  the classes `1 Laser' and `2 Lasers' are predicted less frequently than others as one can read from the \textit{Recall} column. We correlate this  outcome with the fact  that for these specific classes  data corresponding to different levels of powers for the Lasers are merged together. This argument is supported by the fact that for the cases where the EDFA is used, the model is  able to distinguish between different levels of power and the prediction outcomes are very good. Finally, the model shows the highest confidence in distinguishing the different levels of photon loss, particularly in the case of 67\% of photon loss, the model makes statistically no mistakes. 
The Macro average  of \textit{F1-score},  that is  the weighted average mean of Precision and Recall, is  at the value $0.89$ and one
 may conclude   that the developed methodology  has the capacity to distinguish each class of simulated impairment with  high certainty.
 The chord

\begin{table}[h!]
    \centering
    \caption{Classification results for 9 classes of conditional}
    \label{tab:mytable2}
    \begin{tabular}{|r|c|c|c|c||c|}
        \hline
      Class $\#$ &  & Precision & Recall & F1-Score & $\#$ Data \\
        \hline
     \textbf{0}  & Normal & 0.90 & 1.00 & 0.95 & 4064\\
        \hline
     \textbf{ 1}  &  1 Laser  & 0.88 & 0.27 & 0.41 & 395 \\
        \hline
     \textbf{2}  &   2 Lasers  & 0.93 & 0.77 & 0.85 & 466\\
        \hline
    \textbf{3}  &    4 Lasers $\&$ EDFA (18 mA) & 0.98 & 1.00 & 0.99 & 392\\
        \hline
     \textbf{4}  &   4 Lasers $\&$ EDFA (21 mA) & 0.96 & 0.99 & 0.98 & 2070\\
        \hline
     \textbf{5}  &   4 Lasers $\&$ EDFA (24 mA) & 1.00 & 0.87 & 0.93& 353\\
        \hline
    \textbf{6}  &   Photon loss 20\% & 0.96 & 1.00 & 0.98 &  1067\\
        \hline
     \textbf{7} &  Photon loss 46\% & 0.99 & 0.93 & 0.96 & 1529\\
        \hline
     \textbf{8}  &   Photon loss 67\% & 1.00 & 1.00 & 1.00 &  1629\\
        \hline \hline
       & & & && \\
        \hline
       & Accuracy &  & & 0.95 & \\
        \hline
      &  Macro Average & 0.96 & 0.87 & 0.89 & \\
        \hline
     
    \end{tabular}
\end{table}

It is worth noting that the xgboost model alone  using all features, achieves an average $0.86$ F1 score, which is lower than the $0.89$ achieved by the $50$-features fed into the  NN. In typical tabular data an opposite trend is usually observed, and we may conclude that for the given time series the combination of xgboost and NN is preferable. We also studied the performance of NN for  different number of $K$ features and we found out that as $K$ is decreasing the performance is also decreasing.  Finally it is important to underline that even though in theory QBER and SKR data are correlated, in practice we observed that the inclusion of both type of time series in the analysis much improves the  results of classification.

\section{Conclusions}

 The events of photon loss and  addition on  quantum signals are limiting factors across all QKD implementations. In this work, we  designed and successfully tested an ML methodology, that after the training phase, permits to 
identify in real-time such fundamental types of impairments. To the best of our knowledge this is the first  work  dedicated to the detection
of anomalous conditions of a QKD link  by disclosing   information  from QBER's time series. 

In this work the methodology is  tested at a basic level  by emulating anomalous conditions for the QKD link in the lab. We expect that the application of this methodology  in a  QKD test-bed will induce further refinements on the parameters and structure of this initial  model. Finally, an advantage of the  developed methodology is its QKD device-agnostic applicability that we plan to exhibit in  future work.

\section*{Acknowledgements}
This work was supported by   the project Hellas QCI co-funded by the European Union under the Digital Europe Programme  grant agreement  No.101091504. A.M.  acknowledges partial support from the European Union’s Horizon
Europe research and innovation program under grant agreement No.101092766 (ALLEGRO Project).

\bibliography{references}

\end{document}